# Ultrahigh Enhancement of Electromagnetic Fields by Exciting Localized with Extended Surface Plasmons

Anran Li,[1] Sivan Isaacs,[2] Ibrahim Abdulhalim,*,[2] Shuzhou Li*,[1]

[1]School of Materials Science and engineering, Nanyang Technological University, 639798 Singapore

[2]Department of Electro-Optic Engineering and the Ilse Katz Institute for Nanoscale Sciences and Technology, Ben Gurion University, Beer Sheva 84105, Israel

**ABSTRACT:** Excitation of localized surface plasmons (LSPs) of metal nanoparticles (NPs) residing on a flat metal film has attracted great attentions recently due to the enhanced electromagnetic (EM) fields found to be higher than the case of NPs on a dielectric substrate. In the present work, it is shown that even much higher enhancement of EM fields is obtained by exciting the LSPs through extended surface plasmons (ESPs) generated at the metallic film surface using the Kretschmann-Raether configuration. We show that the largest EM field enhancement and the highest surface-enhanced fluorescence intensity are obtained when the incidence angle is the ESP resonance angle of the underlying metal film. The finite-difference time-domain simulations indicate that excitation of LSPs using ESPs can generate 1-3 orders higher EM field intensity than direct excitation of the LSPs using incidence from free space. The ultrahigh enhancement is attributed to the strong confinement of the ESP waves in the vertical direction. The drastically intensified EM fields are significant for highly-sensitive refractive index sensing, surface-enhanced spectroscopies, and enhancing the efficiency of optoelectronic devices.

**KEYWORDS:** localized surface plasmon resonance, surface palsmon polariton, Kretschmann-Raether, metal nanoparticles, electromagnetic field enhancement, surface-enhanced spectroscopies

## 1. INTRODUCTION

Noble metal nanostructures have been widely applied in sensitive refractive index sensing, surface-enhanced Raman scattering (SERS), surface-enhanced fluorescence (SEF), surface-enhanced infrared absorption (SEIRA), and enhancing the efficiency of optoelectronic devices.[1-8] These applications of metallic nanostructures are attributed to their ability to generate intense electromagnetic (EM) fields in the nanoscale vicinity of their surface stemming from the surface plasmon resonances (SPRs). SPR is the coherent oscillation of free electrons at the metal surface with a frequency in resonance with the incident light.[3] Generally, there are two types of SPRs for noble metal nanostructures: propagating (or extended) surface plasmons (ESPs), also known as surface plasmon polaritons (SPPs) and localized surface plasmon resonances (LSPRs).[3, 9] ESPs are usually supported by structures with at least in one dimension approaching the excitation wavelength, for example, thin metal films. ESPs travel along the metal-dielectric interface and the fields of ESPs exponentially decay into both bounding media.[3] In the case of LSPRs, plasmons are sustained by nanoparticles (NPs) much smaller than the incidence wavelength. LSPRs of noble metal NPs give rise to intense absorption, scattering, and extremely enhanced EM fields at their resonance wavelengths.[10]

In the past decades, extensive research work has been done to explore metallic nanostructures that could generate hot spots with large EM field enhancement for high-performance sensors.[11-20] Particularly, the system composed of metal NPs positioned over a continuum metallic film has received increasing attention.[20-28] In metal NP over the metal film (MNOMF) configuration, the NP couples with its mirror image in the metallic film, generating strongly enhanced EM fields localized at the junction between the NP and the metal film.[24, 29-31] The MNOMF can be fabricated using low-cost and simple methods over large areas by directly depositing the NPs on top of the metal film.[20, 24, 32] In addition, separation distance between the NP and the film can be well-tuned by adjusting the thickness of the dielectric spacer, resulting in both the well-tuned resonance position of the supported NP and the highly uniform and reproducible hot spots.[23-25, 31-36] More importantly, MNOMF provides coupling in the vertical direction, for which uniformly distributed gaps between the NP and the film with sub-1 nm size can be fabricated. Such an extremely small gap is critical for generating extremely enhanced EM fields in coupled nanostructures.[25, 35-36] All these advantages make the MNOMF a good platform for SERS, SEF, plasmon-enhanced photoluminescence, and other techniques.[22, 37-40]

Generally, the metal NP and the metal film can sustain LSPRs and ESPs, respectively. For periodic NP array on metallic film, the NP array can excite the ESPs of the metal film by acting as a 2D grating to provide additional momentum. The generated ESPs at the metal-dielectric interface can then couple with the LSPRs or the Bragg waves of the NP array, generating higher EM field enhancement around the NPs.[26-27] However, the ESPs of the underlying metal film in MNOMF were seldom considered in previous studies, because ESPs cannot be directly excited using the free space incidence due to the momentum mismatch.[41]

In the present study, taking the Au nanosphere over the Ag film as an example, we propose a novel methodology to excite the LSP of the supported NPs using the ESP waves of the underlying metallic film. We demonstrate that the excitation of LSPRs using ESP waves can generate much higher EM field intensity than direct excitation of the LSPRs using incidence from free space, attributed to the strong confinement of the ESP waves in the vertical direction. Using finite-difference time-domain (FDTD) method, we demonstrate that the largest EM field enhancement is generated when the incidence angle is the ESP resonance angle of the underlying metal film. In addition, EM field enhancement depends strongly on the Si spacer thickness when exciting the LSPs of the NP using ESP waves, showing drastic decrease when gradually increasing the Si spacer thickness. Considering the match between the LSPRs and the ESPs, the effect of the metal NP size on the EM field enhancement is also examined. Finally, the drastically enhanced EM fields generated by exciting the LSPRs using ESPs are demonstrated in SEF experiments.

## 2. THEORETICAL METHODS

**A. ESP resonance angle of a metallic film.** On a flat metal-dielectric interface, the frequency-dependent wave vector of ESPs is,

$$k_{sp} = k_0 \sqrt{\frac{\varepsilon_m \varepsilon_d}{\varepsilon_m + \varepsilon_d}} \qquad (1)$$

where $k_0 = \omega/c$ is the free-space wave vector. $\omega$ is the frequency of the incident light, $c$ is the velocity of light in vacuum. $\varepsilon_m$ and $\varepsilon_d$ are the complex dielectric constants of the metal and the dielectric medium (analyte), respectively. Equation (1) shows that the momentum of ESPs, $k_{sp}$ is larger than the momentum of a free-space photon $k_0$ with the same frequency. Due to the momentum mismatch, ESPs of a thin metallic film cannot be directly excited by free space incidence.[41] In the present study, we deposit the metallic film on a prism in Kretschmann-Raether (KR) configuration to bridge the momentum mismatch.

In prism-based KR configuration, a transverse magnetic (TM) polarized light is launched from the prism side. The wave vector along the interface is:

$$k_x = 2\pi n_p \sin\theta_p / \lambda = k_0 n_p \sin\theta_p \qquad (2)$$

where $k_0 = \omega/c = 2\pi/\lambda$, $n_p$ is the refractive index of prism, $\theta_p$ is the propagation angle in the prism. In order to generate ESPs on the metal-dielectric interface, it is compulsory to bridge the momentum mismatch between the incident light and ESPs. Hence the condition:

$$\sqrt{\varepsilon_p} \sin \theta_p = Re\left\{\sqrt{\frac{\varepsilon_m \varepsilon_d}{\varepsilon_m + \varepsilon_d}}\right\} \qquad (3)$$

Here, $\varepsilon_p = n_p^2$ is the dielectric constant of the prism. Equation (3) can be used to calculate the ESP resonance angle of the metal film. However, when a dielectric layer (for example, Si) is added on top of the metal film, equation (3) cannot accurately give the resonance angle. In this case, the Abeles 2x2 transfer matrix approach is used to calculate the ESP resonance angle in this article. When the incidence angle is the ESP resonance angle of the metallic film, ESP waves can be generated at the metal-dielectric interface. The generated ESP waves can then hit the supported NP and excite its LSPRs.

**B. Numerical method by FDTD.** Schematic illustration of the setup used to excite LSPRs of a metal NP with ESPs of the metallic film is shown in Figure 1a. A Au NP is deposited on the Ag film separated by a thin Si spacer layer. The thin Ag film is deposited on a SF-11 prism in the KR configuration. A TM polarized light with incidence angle θ is launched from the prism side to gain momentum in the surface direction in order to excite the ESPs at the Ag film surface. The optical properties of the nanosystem shown in Figure 1a are first investigated using 3D FDTD method.[42] In FDTD calculations, the thickness of the Ag film is adjusted to maintain the highest contrast in the reflection spectrum, i.e., keeping the minimum reflection to be zero at the ESP resonance angle. Among all modeling, a total-field scattered-field plane wave source is chosen to estimate the interaction between the propagating plane wave and the metallic nanostructures. Surrounding medium is set to be water since biological and chemical sensing is more popular in water environment. The dielectric function of Au is obtained from Johnson and Christy.[43] The dielectric function of Ag film is achieved from reference 44[44], and the refractive index of Si is obtained from reference 45[45]. To get accurate results, override mesh region with 0.5 nm mesh size is used at the gap region between Au sphere and Ag film, and override mesh regions with 1 nm size are used to cover the Au sphere and Ag film. Before all the simulations, convergence testing was carefully done to verify the accuracy and stability of the calculations. In this study, all the calculations are performed using the FDTD simulation program (FDTD solutions 8.6, Lumerical solutions, Inc., Vancouver, Canada).

## 3. RESULTS AND DISCUSSIONS

The effects of ESP waves on the EM field enhancement in the Au sphere over the Ag film shown in Figure 1a are examined by carefully adjusting the incidence angle to approach the ESP resonance angle of the Ag film. Here, the diameter of the Au nanosphere is 40 nm. The thickness of the Ag film is 46.4 nm. The Au nanosphere and Ag film are separated by 2 nm Si spacer. And the wavelength of the incidence is 655 nm. In this case, the ESP resonance angle of the Ag film is $53.775^\circ$. As shown in Figure 1b, EM fields are dominantly confined at the gap region between the Au nanosphere and the Ag film when the incidence angle is away from the ESP resonance angle. The strongly confined and enhanced EM fields at the junction between the metal NP and metallic film at off resonance angles have been demonstrated before and can be explained by the near-field coupling between the NP and its mirror image in the metallic film.[29-30, 34, 46] When gradually adjusting the incidence angle to approach the ESP resonance angle ($53.775^\circ$) of the Ag film, the EM fields within the gap region become more and more intensely enhanced. At the same time, hot spots with large EM field enhancement are distributed not only at the gap region between the Au sphere and the Ag film, but also at the top of the Au sphere, it looks like "igniting the NP".

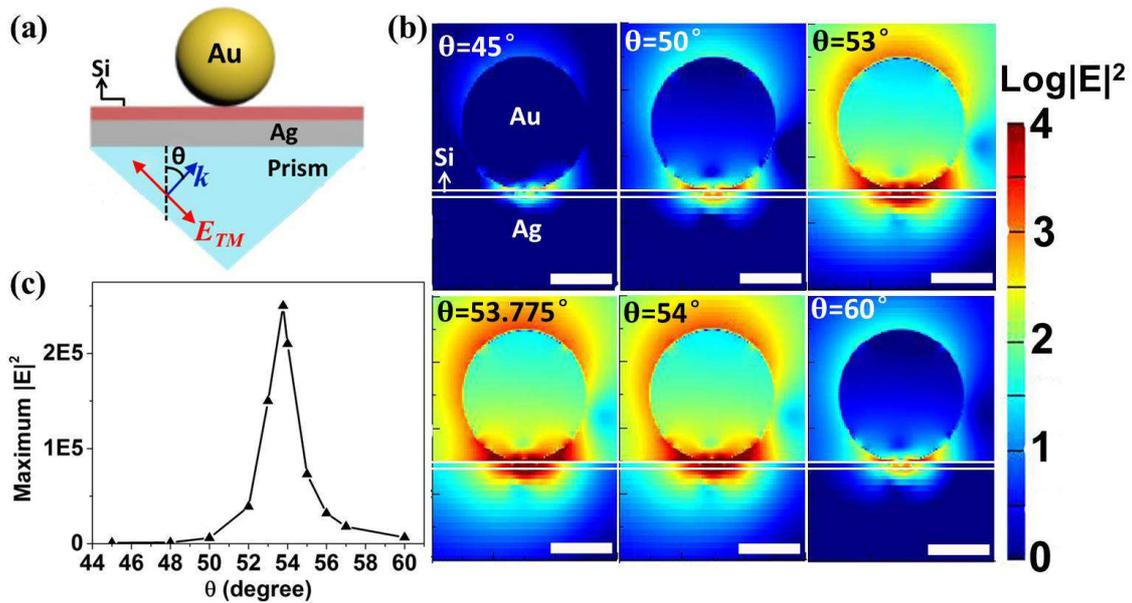

**Figure 1.** (a) Schematic illustration of a Au nanosphere over a Ag film separated by a thin Si spacer. The Ag film is coated on a SF-11 prism in the KR configuration. θ is the incidence angle inside the prism. (b) EM field distributions and (c) maximum EM field enhancement of a 40 nm Au sphere on 46.4 nm Ag film with 2 nm Si spacer at 655 nm with different incidence angles θ. The Ag film is coated on a SF-11 prism. Surrounding medium is water. The scale bar in each figure of (b) is 20 nm.

The maximum EM field enhancement at the Au nanosphere over the Ag film configuration with various incidence angles is calculated and plotted in Figure 1c. The results clearly show that EM field enhancement ($|E|^2/|E_0|^2$) as high as $2.5 \times 10^5$ is generated when the incidence angle is the ESP resonance angle of the Ag film at 53.775°. Small deviation of the incidence angle away from the ESP resonance angle leads to a drastic decrease of the EM field enhancement. The full width at half maximum of the resonance response in Figure 1b is ~1.87°. When the incidence angle is decreased to 45°, the maximum EM field enhancement is only 994, becoming more than 250 times smaller than the value at the resonance angle. In previous investigations, incidence angle of ~60° was considered as the angle that can generate the largest field enhancement in MNOMF.[24, 33] However, according to our calculations shown here, the maximum EM field enhancement for 60° incidence angle is only 6.6x10$^3$, being ~38 times smaller than the value obtained at the ESP resonance angle of Ag film. The EM field enhancement of 2.5x10$^5$ obtained through exciting LSPRs using ESP waves shown here is much larger than that achieved in previous studies, where the maximum EM field enhancement for metal NP over glass substrate with 2 nm separation is typically ~10$^2$, and the maximum EM field enhancement for metal NP over metal film with 2 nm gap at off resonance conditions is typically ~10$^3$ or ~10$^4$.[33, 47]

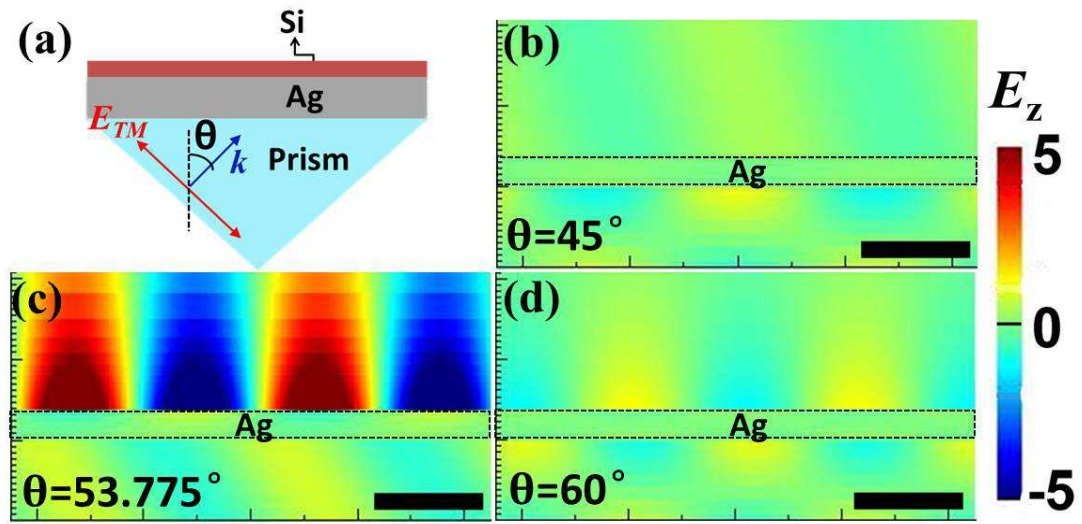

**Figure 2.** (a) Schematic illustration of a 46.4 nm Ag film deposited on a SF-11 prism in KR configuration. The Ag film is covered with 2 nm Si layer. (b~d) $E_z$ distributions of the structure shown in (a) with 45°, 53.775°, and 60° incidence angles. The color bar applies to all the EM field maps. The scale bar in each figure is 200 nm.

The greatly enhanced EM fields in the Au nanosphere over the Ag film at 53.775° are stemming from the generation of the ESP waves at the top surface of the Ag film, as demonstrated by Figure 2. Figure 2 shows the *z* component of the EM field ($E_z$) of a 46.4 nm Ag film deposited on a SF-11 prism with 45º, 53.775º, and 60º incidence angles, respectively. Due to the momentum mismatch, ESP waves cannot be generated at the top surface of the Ag film when the incidence angle is 45º or 60º (Figure 2b, d). When the incidence angle is the ESP resonance angle of the Ag film at 53.775º, the momentum mismatch between the ESPs and the free space incidence is bridged. In this case, the ESP waves are efficiently generated at the top surface of the Ag film (Figure 2c). The generated ESP waves can then hit the Au nanosphere. Figure 2c indicates that ESP waves are strongly confined in the vertical direction. Therefore, the ESP waves can more effectively excite the Au nanosphere, generating much larger EM field enhancement at both the junction between the Au nanosphere and the Ag film and the top surface of the Au nanosphere.

For comparison purpose, the EM field distribution of a 40 nm Au nanosphere deposited on a free-standing Ag film at 655 nm with 53.775º incidence angle is calculated (Figure S1). When moving away the SF-11 prism, ESP waves cannot be excited at the top surface of the Ag film due to the momentum mismatch between the ESPs and the free space incidence (Figure S1a). In this case, the LSPRs of the Au nanosphere are directly excited by the incidence that travels through the Ag film. Figure S1b shows that the maximum EM field enhancement in Au nanosphere over a free-standing Ag film with 53.775º incidence angle is only 361, being ~693 times smaller than the value achieved from exciting the Au nanosphere using the ESP waves. The results demonstrate that excitation of LSPRs using ESP waves generates much larger EM field enhancement than direct excitation of LSPRs using incidence from free space.

When exciting the LSPRs of the metal sphere using the ESP waves of the underlying metallic film, the dielectric spacer between the NP and the film plays a significant role on the EM field enhancement. Figure 3a shows the EM field distributions of a 40 nm Au nanosphere over the 46 nm thick Ag film with various Si spacer thicknesses at 655 nm. The incidence angle for each case is the ESP resonance angle of the Ag film. The ESP resonance angles of the Ag film with 0 nm, 2 nm, 4 nm, 6 nm, 8 nm, and 10 nm thick Si spacers are $52.5^o$, $53.775^o$, $55.525^o$, $58^o$, $61.76^o$, and $67.9^o$, respectively. With increasing the Si layer thicknesses, the field distribution pattern becomes more and more symmetric. At the same time, EM fields at both the nanosphere-film gap region and the top surface of the Au sphere exhibit obvious decrease with gradually increasing Si thicknesses. As shown in Figure 3b, the maximum EM field enhancement is obviously decreased from $2.5 \times 10^5$ to $2.1 \times 10^4$ when increasing the Si thickness from 2 nm to 10 nm. The results shown in Figure 3 indicate that the best EM field enhancement is obtained in the Au sphere over the Ag film without the Si spacer layer. However, a dielectric spacer layer is in fact useful for practical applications. For example, the dielectric spacer layer can be used in purpose to functionalize the surface for specific sensing. In addition, the dielectric spacer layer can be used to control the resonance position of the MNOMF by adjusting its thickness.[34-35] Adding Si layer on top of the metal film layer in the KR configuration has also been demonstrated to narrow the resonance dip in the reflection spectrum and increase both the sensitivity and figure of merit of refractive index sensing.[48]

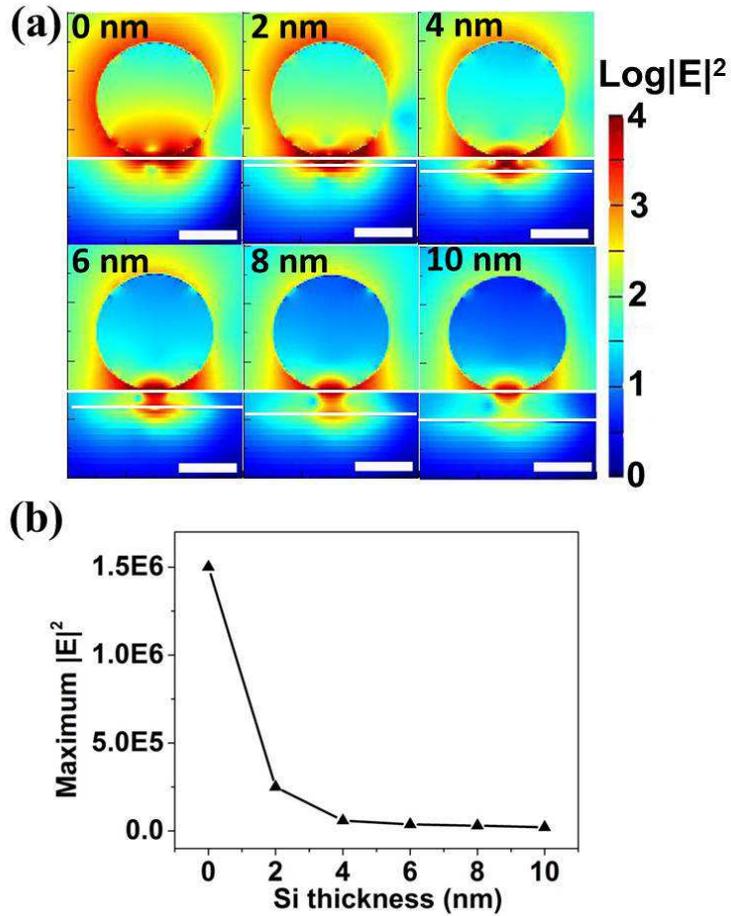

**Figure 3.** EM field distributions (a) and maximum EM field enhancement (b) for a 40 nm Au nanosphere over a 46.4 nm Ag film at 655 nm with various Si spacer thicknesses. The Ag film is coated on a SF-11 prism. The incidence angle is the ESP resonance angles of the Ag film with different Si spacer thicknesses. The scale bar in each figure of (a) is 20 nm.

In above-mentioned calculations, the diameter of the Au nanosphere is randomly fixed at 40 nm without considering the match of the LSPRs of the Au nanosphere and the ESPs of the underlying Ag film. Therefore, extra calculations are performed to check what happens when gradually varying the diameter of the Au sphere (Figure 4). Besides calculations at 655 nm wavelength, additional calculations at 532 nm and 785 nm excitations are carried out since 532 nm and 785 nm wavelengths are usually used in SEF and SERS experiments. The Au sphere and the Ag film are separated by a 2 nm Si spacer. The resonance angles of the Ag film at 532 nm, 655 nm, and 785 nm are 58.7°, 53.775°, and 52.15°, respectively. As shown in Figure 4a, the maximum EM field enhancement is always located at the gap region between the Au nanosphere and the Ag film, regardless of the incidence wavelength and the Au nanosphere diameter. In addition, the EM field enhancement at the top surface of the Au nanosphere decreases with increasing Au nanosphere diameter. Figure 4b shows that EM field enhancement larger than $1 \times 10^5$ can be obtained at 655 nm and 785 nm by adjusting the Au sphere diameter. However, we notice that the maximum EM field enhancement at 532 nm is only ~$10^3$, being much smaller than that at 655 nm and 785 nm. The smaller EM field enhancement at 532 nm is associated with the lower quality factor of Au at 532 nm.[9] Figure 4b indicates that the optimum diameter of Au sphere for EM field enhancement at 532 nm is ~40 nm. For 655 nm and 785 nm excitations, the optimum size of Au nanosphere is 60 nm and 100 nm, respectively. The increase of the optimum size of the Au sphere with increasing incidence wavelength is reasonable, considering the match of the LSPRs of the Au nanosphere and the ESPs of the Ag film. The LSPR position of Au sphere red shifts with rising diameters.[20] Results shown in Figure 4b also indicate that the maximum EM field enhancement in Au sphere over Ag film system does not depend significantly on the size of the Au nanosphere when the sphere diameter is larger than 100 nm.

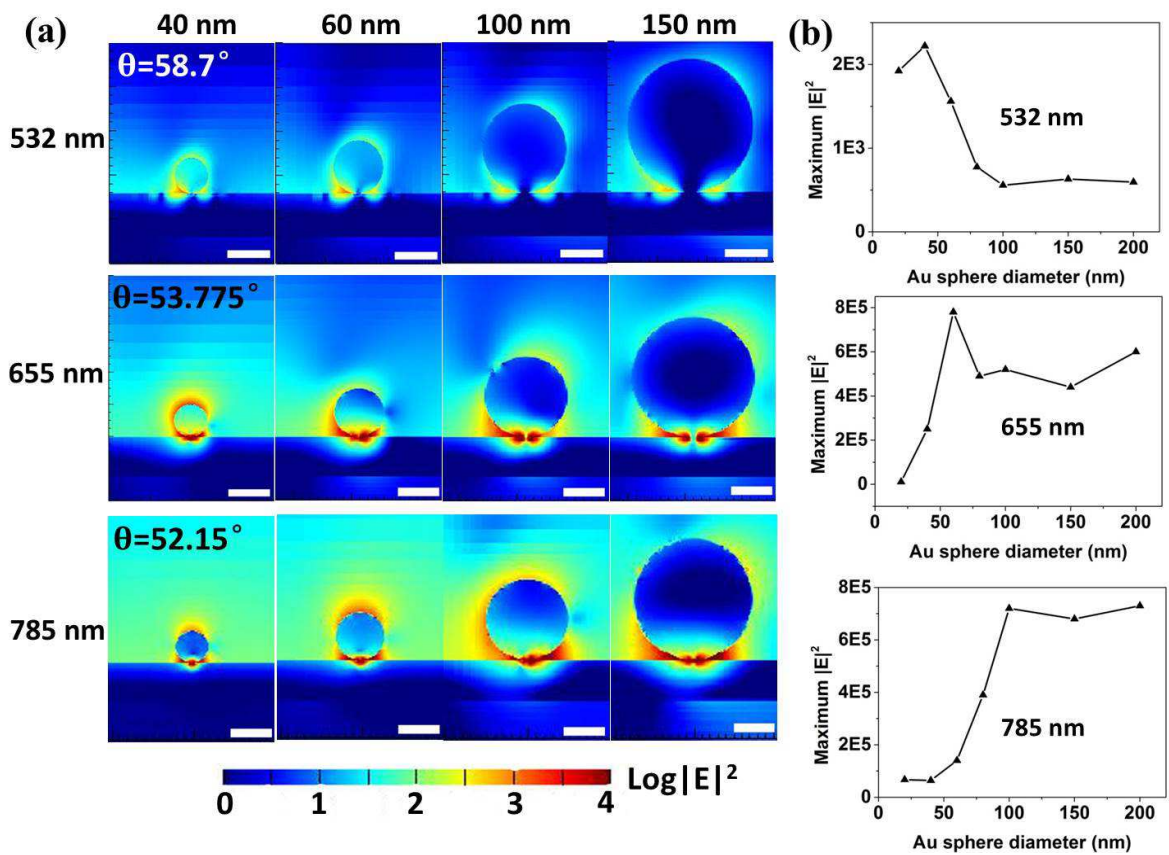

**Figure 4.** (a) EM field distributions for the Au sphere with various diameters over Ag film separated by a 2 nm Si spacer at 532 nm, 655 nm, and 785 nm, respectively. The scale bar is 50 nm. (b) Maximum EM field enhancement as a function of Au sphere diameter at 532 nm, 655 nm, and 785 nm. The Ag film is coated on SF-11 prism. The incidence angles are the ESP resonance angles of Ag film at 532 nm, 655 nm, and 785 nm excitations, respectively.

FDTD calculations indicate that much higher EM field enhancement is obtained in the Au sphere over the Ag film when the ESP waves of the Ag film are excited. These drastically intensified EM fields are extremely essential for highly sensitive surface-enhanced spectroscopies (SES). In an attempt to experimentally prove the theoretical prediction, SEF experiments are performed in this study. Figure 5a shows the fluorescence intensity of a monolayer Rhodamine 6G covered on the Ag film with 532 nm excitation (green laser). The experiments are performed in air. The thickness of the Ag film is 46~47 nm. The Ag film is deposited on a SF-11 substrate which is attached to the SF-11 prism in KR configuration using index matching liquid. Using the Abeles transfer matrix approach, the ESP resonance angle of the Ag film in this case is 37.25$^o$. Fluorescence intensity at two incidence angles of 37.25$^o$ and 60$^o$ is measured, corresponding to the on resonance condition and off resonance condition, respectively. The fluorescence intensity of the monolayer Rhodamine 6G covered on the Ag film without the Au nanosphere deposition is first measured. For the off resonance condition at 60$^o$, the fluorescence intensity of Rhodamine 6G is extremely small with a value of ~0.7 (arbitrary units). When the incidence angle is the ESP resonance angle of the Ag film at 37.25$^o$, fluorescence intensity of Rhodamine 6G is increased to ~3. The enhanced fluorescence intensity from Ag film at 37.25$^o$ is attributed to the enhanced EM fields at the Ag-air interface stemming from the ESPs of the Ag film, a phenomenon known as surface plasmon coupled emission (SPCE). When further depositing 40 nm Au NPs on top of the Ag film, increased fluorescence intensity is observed at both the off resonance and the on resonance conditions. At off resonance condition, the increased fluorescence intensity after depositing Au NPs is associated with the enhanced EM fields at the junction between the NP and the film, originating from the near-filed coupling between the Au NP and its mirror image in the Ag film layer.[29-30, 34, 46] When changing the incidence angle from 60$^o$ to the ESP resonance angle at 37.25$^o$, ~24 times enhanced fluorescence intensity from Au sphere over Ag film configuration is observed. At the on resonance condition with 37.25$^o$ incidence angle, the fluorescence intensity from Au sphere over Ag film configuration is as larger as ~71. The extremely enhanced fluorescence intensity from Au sphere over Ag film at on resonance condition is because of the drastically intensified EM fields at the gap region between the Au sphere and the Ag film. As shown in Figure 5b, much larger EM field enhancement is generated at the junction between the Au nanosphere and the Ag film at 37.25° than that at 60°. According to the calculations, the maximum EM field enhancement at 37.25° is $2.2 \times 10^4$, while the maximum EM field enhancement at 60° is only 289.

The EM field enhancement of the same system at the incidence angle of 15° is also calculated (Figure S2). The maximum EM field enhancement at the off resonance angles of 15° is 262, also being much smaller than that at the ESP resonance angle of 37.25°. Both the FDTD simulation results and the SEF experiments demonstrate that excitation of LSPRs using ESP waves can generate much larger EM field enhancement than direct excitation of the LSPRs using incidence from free space, due to the strong confinement of the ESP waves in the vertical direction. These extremely enhanced EM fields can be harnessed to enhance the optical signals emitted from the molecules located at the neighborhood of the NPs, exhibiting great potentials for high-performance SES, such as SEF, SERS, SEIRA, and other spectroscopic techniques.

In the above-mentioned SEF experiments, the Au NPs are diluted with PBST with a concentration of 25%. SEF experiments using Au NPs with various concentrations are also performed (Figure S3). Regardless of the Au NP concentrations, fluorescence intensity of Au sphere over Ag film system at 37.25° is always much larger than that at 60°. In addition, we find that the 25% concentration of Au NPs gives the largest fluorescence intensity, which may be caused by the balance between the EM field enhancement and the number of the hot spots; however further investigations are needed to fully understand the reason for the optimum enhancement of SEF using Au NPs with 25% concentration.

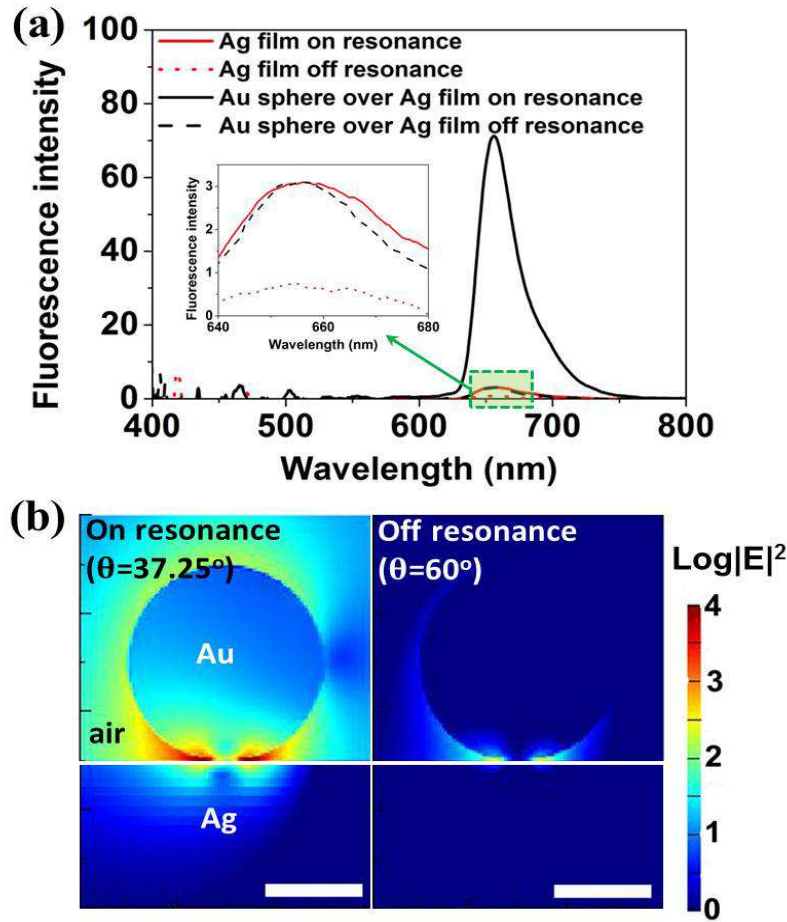

**Figure 5.** (a) Fluorescence intensity of a monolayer of Rhodamine 6G covered on a ~46 nm Ag film with and without 40 nm Au nanosphere deposition at 532 nm. The incidence angles for the on and off resonance conditions are 37.25° and 60°, respectively. The low intensity signals between 640 nm and 680 nm are zoomed in and shown in the inset. (b) Calculated EM field distributions of a 40 nm Au sphere over the 46 nm Ag film at 532 nm with 37.25° and 60° incidence angles. The scale bar in each figure is 20 nm.

**4. CONCLUSIONS**

In conclusion, a novel methodology is proposed to obtain drastically enhanced EM fields in MNOMF through exciting the LSPRs of the supported NP using the ESP waves generated on the metallic film in KR configuration. Due to the strong confinement of the ESP waves in the vertical direction, ESP waves can more efficiently excite the metal NP and hence generate much higher EM field enhancement than free space incidence. We show that the largest EM field enhancement and the highest fluorescence intensity for the Au sphere over the Ag film configuration are obtained when the incidence angle is the ESP resonance angle of the Ag film. In addition, the EM field enhancement is highly dependent on the Si spacer thickness when the ESP waves of the Ag film are excited, where drastic decrease of the EM field intensity with gradually increasing Si spacer thicknesses is observed. It is also shown that the optimum size of the Au sphere for EM field enhancement increases with increasing excitation wavelength, due to the better match between the LSPRs and the ESPs. When the Au sphere and Ag film are separated by 2 nm Si spacer with water surrounding environment, the optimum Au sphere diameters for 532 nm, 655 nm, and 785 nm excitations are 40 nm, 60 nm, and 100 nm, respectively. The drastically enhanced EM fields generated from exciting the LSPRs of NP using ESP waves possess great potentials for sensitive refractive index sensing, high-performance SES, and efficient optoelectronic devices such as better solar cells for energy harvesting.

## ASSOCIATED CONTENT

**Supporting Information**. Additional information provided. This material is available free of charge via the Internet at http://pubs.acs.org.

## AUTHOR INFORMATION

### Corresponding Author

*E-mail: LISZ@ntu.edu.sg and abdulhlm@bgu.ac.il### Author Contributions

The manuscript was written through contributions of all authors. All authors have given approval to the final version of the manuscript.

**Notes**

The authors declare no competing financial interest.

**ACKNOWLEDGMENT**

This Research is conducted by NTU-HUJ-BGU Nanomaterials for Energy and Water Management Programme under the Campus for Research Excellence and Technological Enterprise (CREATE), that is supported by the National Research Foundation, Prime Minister's Office, Singapore.

**TOC GRAPHIC**

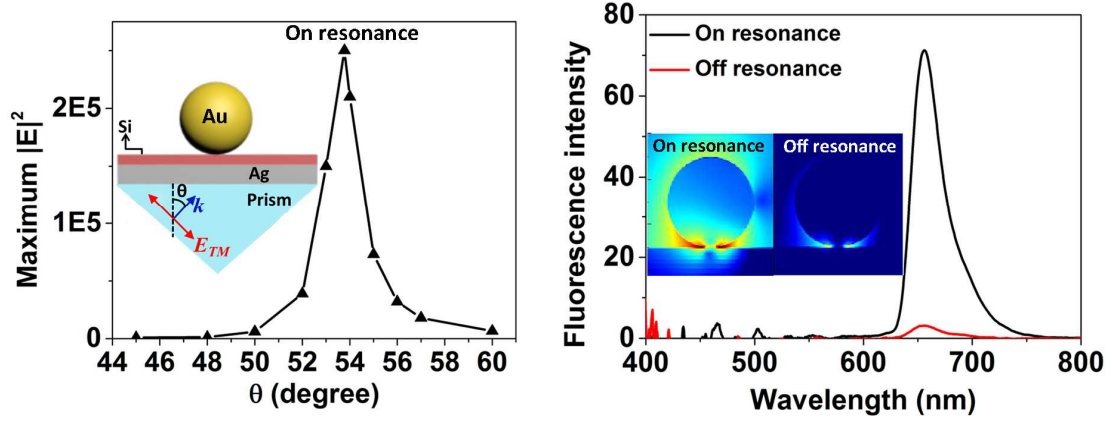

# Supporting Information

**Ultrahigh Enhancement of Electromagnetic Fields by Exciting Localized with Extended Surface Plasmons**

Anran Li,[1] Sivan Isaacs,[2] Ibrahim Abdulhalim,*,[2] Shuzhou Li*,[1]

[1]School of Materials Science and engineering, Nanyang Technological University, 639798 Singapore

[2]Department of Electro-Optic Engineering and the Ilse Katz Institute for Nanoscale Sciences and Technology, Ben Gurion University, Beer Sheva 84105, Israel

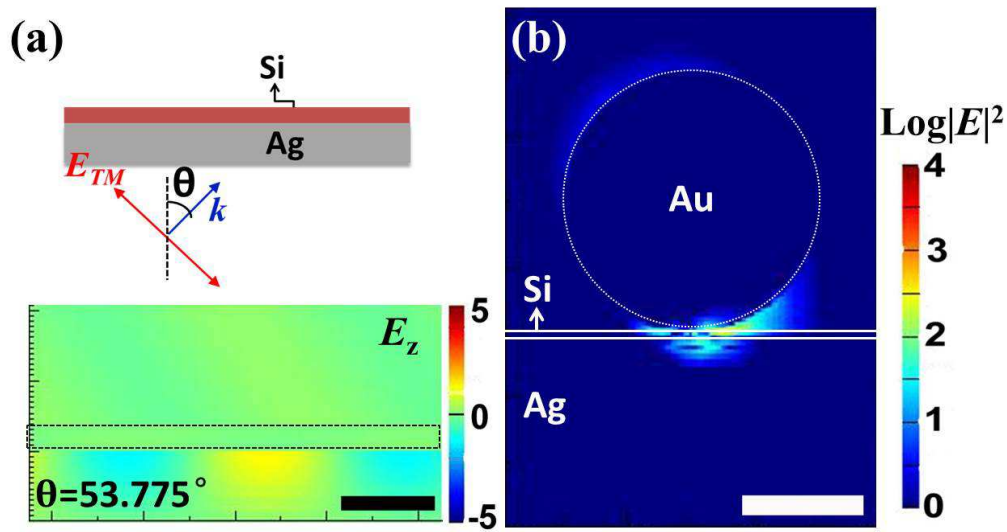

**Figure S1.** (a) $E_z$ distribution of a free-standing Ag film covered with 2 nm Si layer at 53.775° incidence angle. (b) EM field distribution of a 40 nm Au nanosphere on a free-standing Ag film separated by 2 nm Si at 655 nm with 53.775° incidence angle. The scale bars in (a) and (b) are 200 nm and 20 nm, respectively.

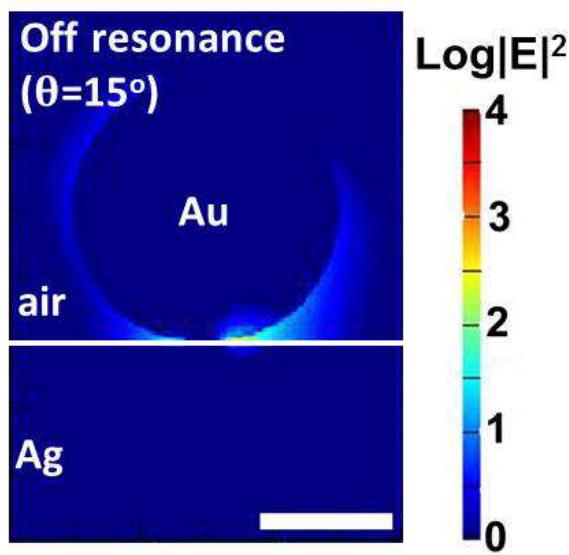

**Figure S2.** EM field distribution of a 40 nm Au nanosphere on top of 46 nm Ag film at 532 nm with 15° incidence angle. The Ag film is deposited on SF-11 prism. The scale bar in each figure is 20 nm.

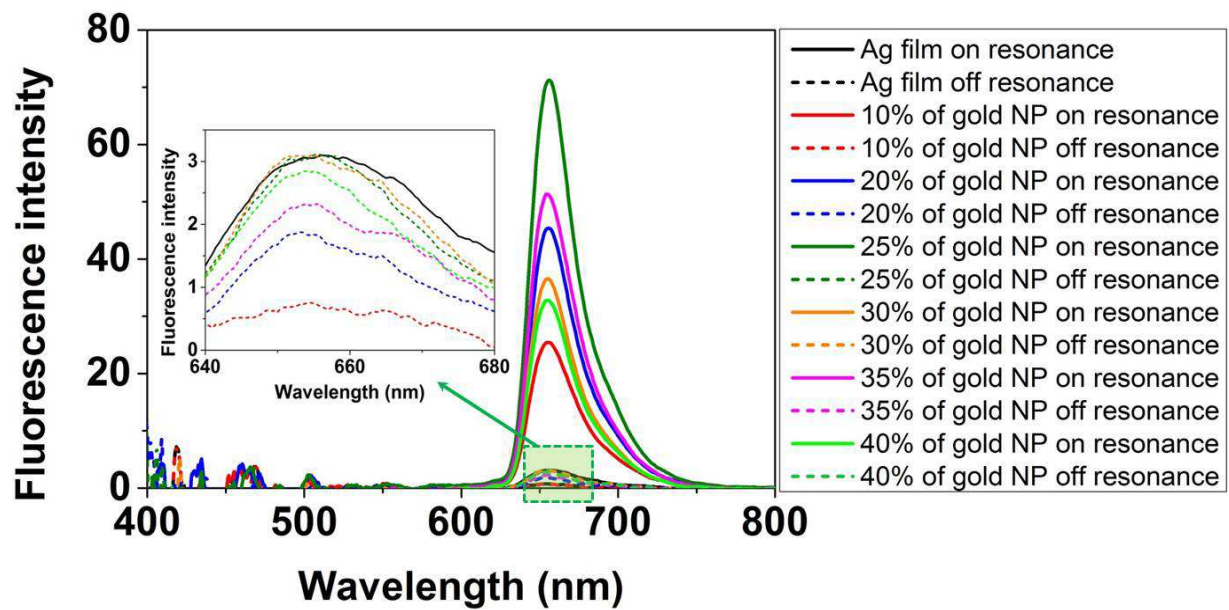

**Figure S3.** Fluorescence intensity of a monolayer of Rhodamine 6G covered on a 47 nm Ag film with various concentrations of 40 nm Au nanospheres deposited on the Ag film at 532 nm. The low intensity signals between 640 nm and 680 nm are zoomed in and shown in the inset. The